\documentclass[12pt,a4paper]{article}
\usepackage[english]{babel}
\usepackage{graphicx}
\usepackage{amsmath}
\usepackage{amssymb}
\usepackage{latexsym}
\usepackage{epsfig}
\usepackage{amscd}

\title{Oscillations of neutrinos produced and detected in crystals}

\author{
A.D. Dolgov\thanks{e-mail: dolgov@itep.ru}\hspace*{2mm}$^{\rm
a,b}$, O.V. Lychkovskiy\thanks{e-mail: lychkovskiy@mail.ru}
\hspace*{2mm}$^{\rm a,c}$, A.A. Mamonov\thanks{e-mail:
mamonov@dgap.mipt.ru.ru} \hspace*{2mm}$^{\rm a,c}$, \\L.B.
Okun\thanks{e-mail: okun@itep.ru}\hspace*{2mm}$^{\rm a}$, M.V.
Rotaev\thanks{e-mail: mrotaev@mail.ru}\hspace*{2mm}$^{\rm a,c}$,
and M.G. Schepkin\thanks{e-mail:
schepkin@itep.ru}\hspace*{2mm}$^{\rm
a}$\\[5mm]
${\rm ^a}$ {\small\it Institute for Theoretical and Experimental
Physics}\\ {\small\it 117218, B.Cheremushkinskaya 25,
Moscow, Russia}\\
${\rm ^b}$ {\small\it INFN, Ferrara 40100, via Paradiso 12,
Italy} \\
${\rm ^c}$ {\small\it Moscow Institute for Physics and
Technology}}

\date{}

\begin{document}

\newcommand{\x}{\mathbf{x}}
\newcommand{\y}{\mathbf{y}}
\newcommand{\p}{\mathbf{p}}
\newcommand{\K}{\mathbf{k}}
\newcommand{\n}{\mathbf{n}}
\newcommand{\q}{\mathbf{q}}
\newcommand{\e}{\mathbf{e}}
\newcommand{\A}{\mathcal{A}}

\maketitle

\abstract{ We analyze neutrino oscillations in a thought
experiment in which neutrinos are produced by electrons on target
nuclei. The neutrinos are detected through charged lepton
production in their collision with nuclei in detector.
Both the target and the detector are assumed to be crystals.
The neutrinos are described by propagators.
We find that different neutrino mass
eigenstates have equal energies.
 We reproduce the standard phase of oscillations and demonstrate
that at large distance from the production point
oscillations disappear.}

\newpage

\section{Introduction}

Neutrino properties are of great interest for particle physics,
astrophysics and cosmology. One of the main sources of information
about neutrinos is an investigation of neutrino oscillations.
Although a great many of papers on neutrino oscillations were
published, there is still no common point of view on this
phenomenon. One way to study neutrino oscillations is to consider
neutrinos in the framework of the standard plane-wave description.
Such an approach neglects effects which concern production and
detection of neutrinos and demands  choosing between two
scenarios:\\ 1) equal momentum scenario (see papers by Gribov and
Pontecorvo \cite{Pontecorvo} and by Fritzsch and Minkovski
 \cite{Fritzsch}),\\
2) equal energy scenario (see papers by
 Lipkin \cite{Lipkin} and Stodolsky \cite{Stodolsky}
 as well as by Kobzarev et al. \cite{Kobzarev} and Grimus and Stockinger
 \cite{Grimus}). In
 ref.\cite{Grimus} neutrino was  created in $\beta$-decay of a neutron
 localized at point $P$ and detected through
 its interaction with an electron localized at point $D$.
 In refs.\cite{Kobzarev} and \cite{DORS} the authors considered
 neutrino produced by an electron beam on the target nucleus $A$ and
 detected through  its interaction with the
 nucleus $B$ of the detector. The whole process looked as (see Fig. 1)
\begin{equation}\label{process}
  e+A+B \rightarrow l+C+D,
\end{equation}
where $l$ is a lepton ($e$, $\mu$ or $\tau$), while $C$ and $D$ are recoil nuclei.
In ref. \cite{DORS} the nuclei $A$ and $B$ were supposed to
be unconfined
in a gaseous target/detector  and  described
by wave packets; the electron wave function was also
assumed to be a wave packet.

In this article we investigate a thought experiment similar to
that considered in ref. \cite{DORS} but with nuclei $A$ and $B$
bound in crystals. We use the rigorous quantum field theory approach
(Feynman diagram with neutrino propagators)  to achieve the
following goals: \\
1) to show that in the case under consideration equal energy
scenario takes place, \\
2) to reproduce the standard form of the oscillation phase, \\
3) to integrate over the phase space of the final particles in order to
obtain the probability of the process,  \\
4) to demonstrate that oscillations disappear at large distances
from the production point,   \\
5) to investigate corrections due to non-zero temperature and
to show that they are not essential for the range of temperatures
at which crystals may exist.

One of the reasons to write this paper is that in the year 2004 in
the most authoritative particle physics review of Particle Data
Group the so-called "equal momentum scenario" was chosen to
describe neutrino oscillations \cite{Kayser}. We think, following
Vysotsky \cite{Vys}, that though this approach gives the standard
result, it is not self-consistent and thus misleading: neutrinos
produced at point $A$ (by electron) would not have a definite
(electronic) flavor, but their flavor would oscillate with time
at point $A$.

In the case of solid state detector with stationary nuclei
the neutrino energy is determined by energy conservation for the
interaction in the detector and does not depend on neutrino
masses. This conforms with the point of view of Lipkin
\cite{Lipkin}. Contrary to that the momentum cannot have a certain
value because of the  uncertainty relation for spatially localized
nuclei.

The paper is organized as follows. In section 2 we derive the
amplitude for the process (\ref{process}) and obtain the
oscillation phase. In section 3 we integrate  modulus of the
amplitude squared over the momenta of the final nuclei and over
the energy of the final lepton. In sections 2 and 3 the
temperature is supposed to be zero. In section 4 the case of
non-zero temperature is considered. Our main results are summarized in section 5.
Appendix contains derivation of factors which suppress oscillations at large distances.

\section{Nuclei bound in crystals at zero temperature}

In this section we follow the lines (and the notations) of
ref.\cite{DORS}. The difference is in the initial wave functions of
the target and detector nuclei: they are stationary in this article
while in ref.\cite{DORS} they were  represented by wave packets.

Consider neutrino production by electron $e$ on a target nucleus
$A$ of mass $M_A$. Then the neutrino collides with nucleus $B$
of mass $M_B$ in a detector and produces charged lepton $l$ (see
Fig. 1) .

The electron neutrino $\nu_e$ produced  on nucleus $A$ is a
superposition of three neutrino mass eigenstates: $\nu_e = \sum_j
U_{e j}\nu_j$, where $\nu_j$ is a state with mass $m_j$, ~$U$ is
a unitary mixing matrix, the first and second indices of which
denote flavor and mass eigenstates, respectively.
The detection of the neutrino by means of its interaction with $B$
results in the projection of three propagating neutrino states
onto the final flavour state $\nu_l= \sum_j U_{l j}\nu_j$.
\begin{figure}
\centerline{\epsfig{file=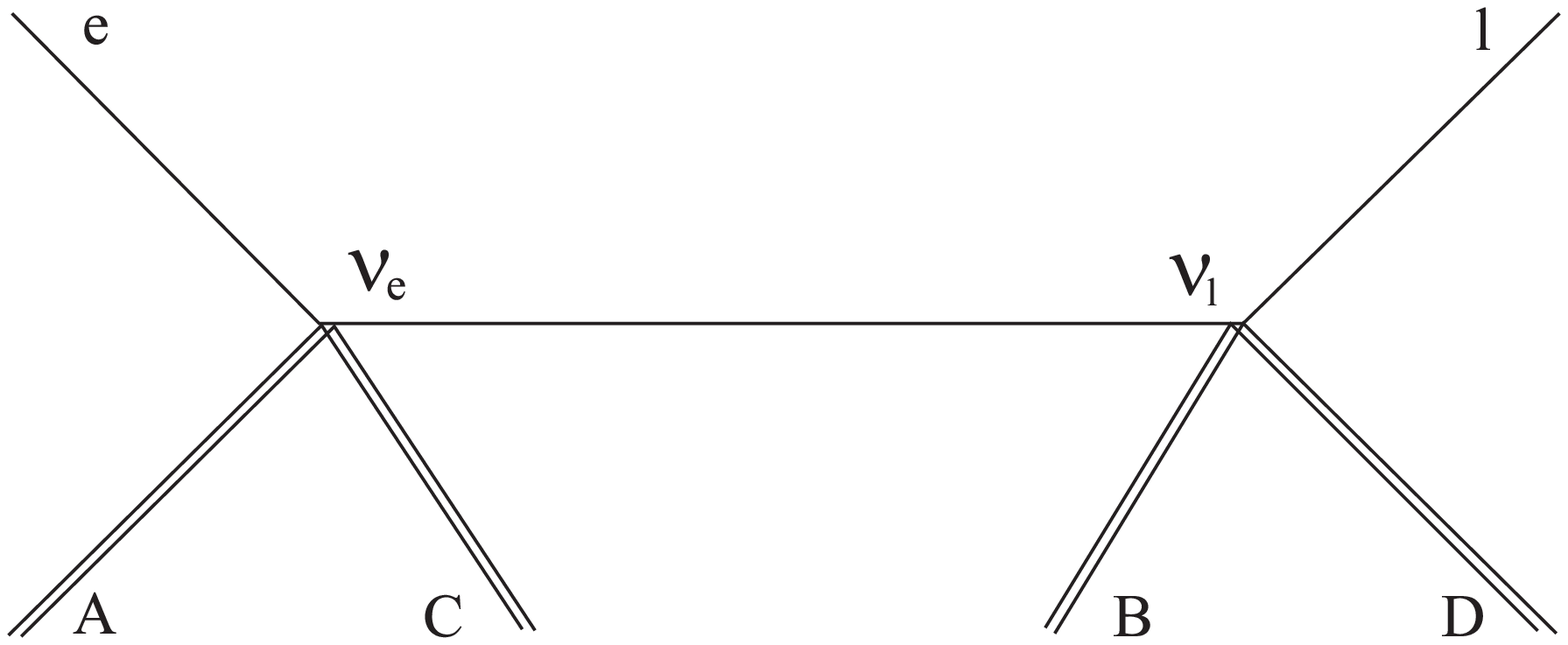, width=10.5cm, height=5cm}}
\caption{Little donkey diagram from ref. \cite{DORS}} \label{}
\end{figure}

The amplitude of the process (\ref{process}) is given by the
following equation:
\begin{equation}
\mathcal{A}_{e\rightarrow l}=\sum_{j}U_{ej}^*U_{lj}\mathcal{A}_j,
\label{A_el}
\end{equation}
where $A_j$ is the amplitude for a given neutrino state of mass
$m_j$ \footnote {The equality signs in equations throughout the
paper should be taken "with a grain of salt" because we omit some
obvious factors, such as coupling constant, $(2\pi)^{-3}$
etc. This makes the formulas easier to read without influencing
the physical results related to oscillations.}:

\begin{equation}
\mathcal{A}_j = \int d^4 x_2 d^4 x_1 \Psi_B (x_2) \Psi^*_l (x_2)
\Psi_D^* (x_2) G_j (x_2-x_1) \Psi_A (x_1) \Psi_e (x_1) \Psi_C^*
(x_1)~.
\label{Aj}
\end{equation}

Here $\Psi_a$ are wave functions of particles $a$; $G_j$ is the
Green function of $j$-th mass eigenstate of neutrino, and
$x_k = (t_k,{\bf x}_k)$ are 4-dimensional coordinates.

There  exists a vast literature in which not only incoming
but also outgoing particles are described by wave packets
 (see review by Beuthe \cite{Beuthe}). We describe the
electron by a wave packet, the initial nuclei $A$ and $B$ -- by
stationary wave functions, while the outgoing particles -- by
plane waves as representatives of a complete set of orthogonal
states.

We assume that the initial nuclei $A$ and $B$ are bound in
crystals and their wave functions are energy eigenstates localized
near central point ${\bf x}_A$ and ${\bf x}_B$, respectively, with
the uncertainty of the order of crystal spacing $a_{A,B}$. Their
wave functions are taken as a product:
\begin{equation}
\Psi_a (x) = \psi_a ({\bf x}-{\bf x}_a) ~ e^{-it E_a}~,
\label{Psia}
\end{equation}
where $a=A,B$ and $E_a$ is the energy of nucleus in the potential of a crystal cell, which is equal to the difference of $M_a$ and binding energy. The Fourier transform of $\psi_a({\bf x})$, which we need in what follows, is

\begin{equation}
K_a({\bf q}_a) = \int d{\bf x} \psi_a ({\bf x}- {\bf x}_a)
e^{-i{\bf q}_a ({\bf x}-\x_a )} \label{K_a}
\end{equation}

By assumption, nuclei $A$ and $B$ are at rest and thus
$K({\bf q}_a)$ is centered
near ${\bf q}_a = 0$ with uncertainty $\sigma_a\sim a_a^{-1}$.

The wave function of the incident electron is taken as a wave packet:
\begin{equation}
\Psi_e (x) = \int d{\bf q}_e K_e ({\bf q}_e -{\bf p}_e)
e^{i{\bf q}_e ({\bf x} - {\bf x}_e) - iE_et},
\label{Psie}
\end{equation}
where $E_e({\bf q}_e) = \sqrt{{\bf q}_e^2 +m_e^2}$, the Fourier
amplitude $K_e({\bf q}_e -{\bf p}_e)$ is centered  near
${\bf q}_e ={\bf p}_e$ with uncertainty
$\sigma_e$, and the maximum of the envelope of the packet
is at the point ${\bf x}_e$ at the moment $t=0$.

According to the measurements of Novosibirsk group (Pinaev et al., \cite{Pinaev1},\cite{Pinaev2}), which used an undulator
at electron storage ring,  $\sigma_e>0.7\times10^{-6}$ eV
= (30 cm)$^{-1}$. The theoretical analysis of this data and the mechanism
of reduction of the wave packet of a relativistic charged particle by
emission of a photon has been performed by Faleev \cite{Faleev}. (For
the general theory of wave packets see lectures by Glauber
\cite{Glauber}.) As for the upper bound on $\sigma_e$, it is smaller
then $10^{-3}E_e$ for an electron accelerator according to the PDG
(\cite{PDG}, pp.239-241).

For the wave functions of the outgoing particles we can take any complete and
orthogonal set of functions, the most convenient would be just plane waves:
\begin{equation}
\Psi_c^* (x) = e^{ i t E_c - i{\bf p}_c {\bf x} },
\label{Psic}
\end{equation}
where $c=C,D,l$.

It is clear that fermionic nature of neutrino (as well as of
$e$ and $l$) is not essential in the problem at high enough
energies, therefore we replace the neutrino Green function by the
Green function of a scalar particle of mass $m_j$, where $j$
enumerates neutrino mass eigenstates, $j=1,2,3$.  Thus:

\begin{equation}
\label{prop} G_j(\x,t)=\frac{1}{(2\pi)^4} \int \frac{e^{-i\omega
t+i\K\x}} {\omega^2-\K^2-m^2_j+i\varepsilon}d\K d\omega~.
\end{equation}

We integrate in (\ref{Aj}) over $(t_1+t_2)$ and $(t_2-t_1)$ using
\begin{equation} \int
e^{i\omega(t_2-t_1)}G_j(\x_2-\x_1,t_2-t_1)d(t_2-t_1 ) = -
\frac{1}{4\pi|\x_1-\x_2|}e^{i\sqrt{\omega^2-m_j^2}|\x_1-\x_2|}
\end{equation}
and obtain

$$
\mathcal{A}_j=-\int d\x_1 d\x_2 d\q_e
K_e(\q_e-\p_e)e^{i\q_e(\x_1-\x_e)} \psi_A(\x_1-\x_A)
\psi_B(\x_2-\x_B)
$$
\begin{equation}
\label{another_A}
 e^{-i\p_C\x_1-i\p_D\x_2-i\p_l\x_2}~
\frac{ e^{ik_j|\x_1-\x_2|}}{4\pi |\x_1-\x_2|}~
\delta(E_e(\q_e)+E_A-E_C-\omega)~,
\end{equation}
where
\begin{equation}
\label{omega} \omega \equiv E_l+E_D-E_B
\end{equation}
and $~k_j\equiv \sqrt{\omega^2-m_j^2}.$ The delta-function in
eq.(\ref{another_A}) corresponding to the energy conservation for the
whole process appears due to integration over $(t_1+t_2)$.

We would like to emphasize that the energy $\omega$ of the
virtual neutrino
does not depend on $j$. The energy $\omega$ is determined by
energy conservation at point $B$. Due to stationarity of nuclei
$B,D$ and lepton $l$ (see eqs.(\ref{Psia}) and (\ref{Psic})) the energy
$\omega$ is the same for all neutrino mass eigenstates.

Note that $e^{ik_j|\x_1-\x_2|}$ is the only factor in
eq.(\ref{another_A}) which depends on $j$ and leads to
mass-dependent effects such as oscillations. Since the initial
nuclei $A$ and $B$ are in the potential wells, they are well
localized near the points $\x_A$ and $\x_B$, respectively;
therefore the integrand in eq.(\ref{another_A}) is essentially
different from zero if $|\x_1-\x_A|\lesssim a_A$ and
$|\x_2-\x_B|\lesssim a_B$. Hence the effective range of
integration over $\x_1$ and $\x_2$ is limited by $\left
||\x_1-\x_2|-|\x_A-\x_B|\right | \lesssim a$, where $a_A\sim a_B
\sim a$. Furthermore, $k_j$ differs slightly from $ \omega $, so
we may expand
\begin{equation}
\label{k_j} k_j=\omega-\frac{m_j^2}{2\omega}+...
\end{equation}

In what follows we introduce the following notation:
\begin{equation}
x_{AB} \equiv |\x_A-\x_B|~. \label{xAB}
\end{equation}
Then
$$e^{ik_j|\x_1-\x_2|}=e^{ik_jx_{AB}}
\exp{\left[i\left(\omega+O\left(\frac{m_j^2}{2\omega}\right)\right)
(|\x_1-\x_2|-x_{AB})\right]}=$$

\begin{equation}\label{exponent evaluation}
=e^{ik_j x_{AB}}e^{i \omega
(|\x_1-\x_2|-x_{AB})} \left[1+O\left(\frac{m_j^2}{2\omega}a \right)\right]~.
\end{equation}

Lattice spacing $a$ is of order of $10^{-8}$ cm. As for the $\omega$
(which is roughly equal to the electron energy), it can vary
depending on the beam energy of the accelerator which produces
electrons. Let us take, for example,$~\omega = 1$ GeV. This energy
is sufficient for muons to be created in the detector. If we take
$3$ eV as the upper bound for neutrino mass we get
$a m_j^2 /(2\omega) < 10^{-12}$, hence the $j$-dependent correction
in brackets
in eq.(\ref{exponent evaluation}) can be neglected, and we obtain
from   eqs.(\ref{another_A}) and (\ref{exponent evaluation}):

$$
\mathcal{A}_j= -e^{ik_jx_{AB}}\int d\x_1 d\x_2
d\q_e K_e(\q_e-\p_e)
e^{i\q_e(\x_1-\x_e)}
e^{-i\p_C\x_1-i\p_D\x_2-i\p_l\x_2}
$$
\begin{equation}
\label{prefinal_A}
\psi_A(\x_1-\x_A)
\psi_B(\x_2-\x_B)
 \frac{e^{i \omega (|\x_1-\x_2|-x_{AB})}}
 {4\pi |\x_1-\x_2|}
\delta(E_e(\q_e)+E_A-E_C-\omega)~,
\end{equation}

We see that $e^{ik_j x_{AB}}$ is the only $j$-dependent
factor in eq.(\ref{prefinal_A}), which means that the phase
difference between $\mathcal{A}_i$ and $\mathcal{A}_j$ has the
following form :
\begin{equation}
\label{ocsillation_phase}
\phi_{ij}\equiv\phi_i-\phi_j=(k_i-k_j)x_{AB}.
\end{equation}

There is no term, containing energy difference, in the right-hand
side of eq.(\ref{ocsillation_phase}). This results from equality
of energies of different neutrino mass eigenstates, emphasized
earlier. The form of the phase in eq.(\ref{ocsillation_phase})
allows to resolve the problem "equal energy \textit{vs} equal
momentum" in favor of equal energy scenario.

From eqs.(\ref{k_j}) and (\ref{ocsillation_phase}) we obtain the
standard expression:
\begin{equation}
\label{standard_phase}
\phi_{ij}=-\frac{\delta m_{ij}^2}{2\omega}x_{AB},
\end{equation}
where  $~\delta m_{ij}^2\equiv m_i^2-m_j^2$.

To integrate over $\x_1$ and $\x_2$ in eq.(\ref{prefinal_A}) we
take into account that  $a_A,a_B\ll x_{AB}$, and hence we can
use the expansion

\begin{equation}
\label{x_1-x_2} |\bf{x}\mit_1-\bf{x}\mit_2|\simeq x_{AB}-
\bf{n} (\x_1-\x_A) +  \bf{n} (\x_2-\x_B)~,
\end{equation}
where
\begin{equation}
\n=\frac{\x_B-\x_A}{x_{AB}}~.
\end{equation}

From eqs.(\ref{x_1-x_2}) and (\ref{prefinal_A}) we obtain
(up to a constant phase factor):
\begin{equation}
A_j = \frac{e^{ik_j x_{AB}}}{x_{AB}}\, \int d{\bf q}_e
e^{i\q_e(\x_A-\x_e)} K_e ({\bf q}_e -{\bf p}_e)
 K_A ({\bf q}_A) K_B ({\bf q}_B)
\delta \left( E_{in} -  E_{fin}\right)
\label{Aj2}
\end{equation}
where
$$
E_{in} = E_A+E_B+E_e~,~~
$$
$$
E_{fin} =E_C+E_D+E_l~,
$$
$$
{\bf q}_A = {\bf p}_C+\omega\n -{\bf q}_e~,
$$
$$
{\bf q}_B = {\bf p}_l+{\bf p}_D-\omega\n~.
$$
 As follows from  eq.(\ref{Aj2}) and the expressions for the wave
functions presented above, the amplitude is a function of the
vector $\n$, neutrino mass, $m_j$, energies of the initial nuclei,
$E_A$ and $E_B$, central momentum of the incoming electron ${\bf p}_e$,
and momenta of the final particles ${\bf p}_C,~\p_D,~\p_l$.

In the next section we will use an explicit formula for the
amplitude $A_j$ (eq.(\ref{Aj2})) in the case of certain simple
expressions for the crystal potential and for the wave packet of
the electron. We assume that the electron is described by
one-dimensional Gaussian wave packet with
definite direction $\e=\p_e/|\p_e|$, that is
\begin{equation}
d\q_e=\delta \left(\frac{\q_e}{|\q_e|}-\e \right)
~|\q_e|^2~ d|\q_e| ~d\Omega_e
\label{dq_e}
\end{equation}
\begin{equation}
K_e(\q_e-\p_e)=\exp\left[-\frac{(\q_e-\p_e)^2}
{2\sigma_e^2} \right].
\label{Electron gauss}
\end{equation}

We also assume that the potential is an oscillator near the
center of the crystal cell and a constant at large distances
from this point. Furthermore we consider the initial nuclei in
the ground states in their crystal cells:
\begin{equation}\label{nucleus a in the crystal}
K_a(\q_a)=\exp\left({-\frac{\q_a^2}{2\sigma_a^2}}\right),
\end{equation}
where $a=A,B.$
After integration over $\q_e$ in eq.(\ref{Aj2})
by using delta-function in the integrand of eq.(\ref{Aj2})
and delta-function in eq.(\ref{dq_e}) we obtain

$$
 A_j=E_e^0q_e^0~
\frac{e^{ik_jx_{AB}}}{4\pi x_{AB}}~ e^{i\q^0_e(\x_A-\x_e)}
$$
\begin{equation}
\label{final A_j}
\exp\left[{-\frac{(q_e^0\bf{e}\mit-\bf{p}\mit_e)^2}{2\sigma_e^2}}\right]
\exp\left[{-\frac{(\omega\bf{n}\mit+\bf{p}\mit_C-q_e^0\bf{e}\mit)^2}
{2\sigma_A^2}}\right]
\exp\left[{-\frac{(\bf{p}\mit_l+\bf{p}\mit_D-\omega\bf{n}\mit)^2}
{2\sigma_B^2}}\right],
\end{equation}
where
\begin{equation}\label{E_e^0}
E_e^0=E_l+E_C+E_D-E_A-E_B,
\end{equation}
 and
\begin{equation}\label{q_e^0}
q_e^0=\sqrt{(E^0_e)^2-m_e^2}.
\end{equation}
Thus introduced $q_e^0$ coincides
with $p_e = |{\bf p}_e|$ within $\sigma_e$.

\section{Integration over phase space}

In this section we integrate the probability of the process
(\ref{process}) over the phase volume of the final nuclei and over
the energy of the final lepton, assuming explicit form of the
crystal potential and of the electron wave packet (see the end of
the previous section). We demonstrate that due to the neutrino
energy dispersion the oscillations are suppressed at large
distances. We think  that this result is valid for
arbitrary shapes of wave packets.

We consider a situation when only the direction of the final
lepton  is measured, while its energy is not measured and final
nuclei are not registered \footnote
{In fact, as it is clear from
calculations presented in the Appendix, if we do measure the energy
of the final lepton, but  the precision of the measurement is worse
then $\sigma_e$, the result does not change essentially.}.
Thus we
are interested in the differential probability for the final
lepton to be detected in the solid angle $d \Omega_l$ :
\begin{equation}\label{differential probability}
\frac{d P_{e\rightarrow l}}{d \Omega_l}\equiv \int
|\mathcal{A}_{e\rightarrow l}|^2 \frac{d\p_C}{2E_C}
\frac{d\p_D}{2E_D} \frac{p^2_ld p_l}{2E_l}= \sum_{ij}
U_{ej}^*U_{lj}U_{ei}U^*_{li}{\cal P}_{i j}.
\end{equation}
Here $p_l\equiv|\p_l|$; $~\p_C$, $\p_D$ and $\p_l$ are the momenta
of the final nuclei and the final lepton, correspondingly, and

\begin{equation}\label{mathcal P}
{\cal P}_{i j}\equiv\int P_{ij} \frac{d\p_C}{2E_C}
\frac{d\p_D}{2E_D} \frac{p^2_ld p_l}{2E_l},
\end{equation}
where
\begin{equation}\label{P_ij}
P_{ij}\equiv \mathcal{A}_i^*\mathcal{A}_j.
\end{equation}

We interpret the electron detection as $\nu_e$ survival, and $\mu$ or
$\tau$ detection as $\nu_{\mu}$ or $\nu_{\tau}$ appearance,
respectively.

From eq.(\ref{final A_j}) we get:

\begin{equation}
\label{Probability}
P_{ij} =
\frac{\exp \left( i\frac{\delta m_{ij}^2}{2\omega}x_{AB} \right)}
{x_{AB}^2}
\exp{\left[ -\frac{(q_e^0{\bf e}-{\bf p}_e)^2}{\sigma_e^2}
-\frac{(\omega\bf{n}\mit+\bf{p}\mit_C-q_e^0\bf{e}\mit)^2}
{\sigma_A^2}
 -\frac{(\bf{p}\mit_l+\bf{p}\mit_D-\omega\bf{n}\mit)^2}
{\sigma_B^2}\right]}.
\end{equation}
Here we delete inessential factors $1/(4\pi)^2$ and $(E_e^0q_e^0)^2$.
In what follows to simplify formulas we take

$M_C=M_D \equiv M \sim 100$ GeV, $\sigma_A=\sigma_B
\equiv \sigma \sim 1$ keV; 10 MeV $\lesssim E_e \lesssim 1$ GeV.

For such choice of parameters the electron is relativistic, while

\begin{equation}\label{inequalities}
\frac{p_e}{M}\ll 1,
\end{equation}
where $p_e\equiv |\p_e|$.

We calculate $\mathcal{P}_{ij}$, defined by eq.(\ref{mathcal P}),
in the Appendix for two cases:\\
1. Large $\sigma_e$:
\begin{equation}\label{Large}
\sigma_e\gg\sigma\frac{p_e}{M},
\end{equation}
2. Small $\sigma_e$:
\begin{equation}\label{small}
\sigma_e\ll\sigma\frac{p_e}{M}.
\end{equation}
We assume that vectors $\e$ and $\n$ have a substantially nonzero
angle  between them and the module of their difference is of order
of unity. We present the results in the form \footnote{The
exponential character of suppression arises from the Gaussian form
of the initial wave functions in $p$-space.}
\begin{equation}\label{P_FIN}
{\cal P}_{i j}\sim  \exp\left( i \frac{x_{AB}}{L^{osc}_{ij}}
\right) \exp\left[-\left(\frac{x_{AB}}{L^{sup}_{ij}}
\right)^2\right]~,
\end{equation}
where $L^{osc}_{ij}\equiv 2 E_{\nu}/ \delta m^2_{i j}$, and, in
our case, $E_{\nu}=p_e(1+O\left(\frac{p_e}{M}\right))$.
\footnote{Quite often $L^{osc}$ is defined as $4\pi E_{\nu}/
\delta m^2$.} For the first case (\ref{Large})
\begin{equation}
L^{sup}_{ij}=L^{osc}_{ij}\frac{2p_e}{\sigma_e}~,
\label{supp1}
\end{equation}
(see eq.(\ref{FIN-FIN})), while for the second case (\ref{small})
\begin{equation}
L^{sup}_{ij} \sim L^{osc}_{ij}\frac{M}{\sigma}, \label{supp2}
\end{equation}
(see eq.(\ref{final})).

The origin of the above suppression is neutrino energy dispersion,
on which the oscillation length depends. A rather lengthy way to
derive eqs.(\ref{P_FIN})-(\ref{supp2}) is given in the Appendix.

A simple qualitative estimate of the suppression, based on consideration
of two particle reaction $e+A \rightarrow \nu+C$, will be presented in
ref.\cite{WaveFunction}.

It is interesting to note that the suppression length for large $\sigma_e$ (the first case) is equal to that which arises due to spatial separation of the neutrino wave packets with different $m_j$, and hence different velocities, in the case when neutrinos are described not by a propagator,
but by a wave function (see  Dolgov et al. \cite{WaveFunction}, Nussinov \cite{Nussinov}, Kayser \cite{KaW} and Dolgov \cite{Dolgov}).

As for the case of vanishingly small $\sigma_e$, there is no
separation of neutrino wave packets considered in
refs.\cite{WaveFunction}-\cite{Dolgov}, but suppression exists
due to virtual neutrino energy dispersion.

Let us end up this section by considering a toy model of two
neutrino flavours ($\nu_e$ and $\nu_{\mu}$, for definiteness).
In this case from eqs.(\ref{differential probability}) and
(\ref{P_FIN}) we easily find
\begin{equation}
\nonumber \frac{d P_{e\rightarrow e}}{d \Omega_e} =\left\{
1-\frac{1}{2}(\sin 2\theta)^2
\left[1-\exp{\left(-\frac{x^2_{AB}}{L^2_{sup}}\right)}
\cos\left(\frac{x_{AB}}{L_{osc}}\right)\right]\right\}
W(\mathbf{l}) ,~
\end{equation}
\begin{equation}\label{probabilities}
\frac{d P_{e\rightarrow \mu}}{d \Omega_{\mu}}= \frac{1}{2}(\sin
2\theta)^2 \left[1-\exp{\left(-\frac{x^2_{AB}}{L^2_{sup}}\right)}
\cos\left(\frac{x_{AB}}{L_{osc}}\right)\right] W(\mathbf{l}) ~,
\end{equation}
where $\mathbf{l}\equiv\frac{\p_l}{p_l},~
U_{e1}=\cos\theta,~U_{e2}=\sin\theta$ and the factor
$W(\mathbf{l})$ does not depend on the lepton flavour $l$ in the
limit of ultra-relativistic final leptons.

\section{Nuclei bound in crystals at non-zero \\ temperature}
Now we show that for a non-zero temperature the results do
not essentially change. In this case we cannot assume nuclei wave
functions in crystal cells to be stationary. A general expression
for the wave functions of the initial nuclei is the following:
\begin{equation}
\label{nucleus wave function of general form}
\Psi_{A,B}(\x,t)=\sum_n C_{A,B}^n \psi_{A,B}^n(\x)~
 e^{-iE_{A,B}^nt},
\end{equation}
where $n$ enumerates the states with definite energies
$E_{A,B}^n$, while $C_{A,B}^n$ is the amplitude of probability to
measure energy $E_{A,B}^n$ in the state with wave function
$\Psi_{A,B}(\x,t)$. Using the linearity of amplitudes with respect
to initial wave functions we obtain
\begin{equation}\label{general P ij}
P_{ij}=\sum_m \sum_{m'} C_m C^*_{m'} \A^*_{i~m'} \A_{j~m},
\end{equation}
where for convenience we define the two-dimensional index $m\equiv
(n_1,n_2)$, $C_m\equiv C_{A}^{n_1}C_{B}^{n_2}$, and $\A_{i~m}$
stands for $\A_{i}$ calculated with $\psi_{A}^{n_1}$ and
$\psi_B^{n_2}$. For typographical reasons there is no difference
between upper and lower indices.

Since we consider the case of a thermal equilibrium,  we have to
average $P_{ij}$ using $\overline{C_m
C^*_{m'}}=\delta_{mm'}e^{-E_m/T}$, where $E_m\equiv
E_A^{n_1}+E_B^{n_2}$. After averaging we obtain

\begin{equation}
\label{averaged P ij}
\overline{P_{ij}}=\sum_m e^{-\frac{E_m}{T}} \A^*_{i~m} \A_{j~m} .
\end{equation}

We see that taking temperature into account results in a simple
averaging of $P_{ij}$  over
different stationary initial states weighted with $e^{-E_m/T}$,
which is a common case in statistical mechanics. Such a
correction does not result in any observable effect due to the
smallness of thermal energy $T$ compared with $\omega$. Indeed,
\begin{equation}
\phi_{ij}=-\frac{\delta m^2_{ij}}{2\omega}x_{AB} +
 O \left(\frac{\delta
m^2_{ij}}{2\omega}\frac{T}{\omega}x_{AB}\right),
\end{equation}
where $\omega$ corresponds to the case of zero temperature (ground
states, $n_2=0$):
\begin{equation}
\omega=E_l+E_D-E_B^0.
\end{equation}
We see that the correction is essential if
$x_{AB}\sim \ (\omega/T)L_{osc}\sim 10^{10}~L_{osc}~$
for $ \omega=
1$ GeV, $T=300$ K. In the previous section it was shown that at
such distances oscillations are washed out. Thus we may safely
neglect this correction and recover to the standard expression
(\ref{standard_phase}).

\section{Conclusions}

1.Our calculations explicitly confirm the equal energy scenario,
advocated by Lipkin and Stodolsky (see text after
eqs.(\ref{omega}) and (\ref{ocsillation_phase})).
\newline
2.The oscillation phase has its standard form, see
eq.(\ref{standard_phase}).
\newline
3. When integrated over the phase volume of the final nuclei and
over the energy of the final lepton, the oscillating term in the
probability of the process under consideration vanishes
exponentially at the distances
$L_{sup} = (2 p_e / \sigma_e)L_{osc}$ for large $\sigma_e$,
and $L_{sup} \sim (M/\sigma)L^{osc}_{ij}$ for small $\sigma_e$
including plane wave limit for electron;
see eqs.(\ref{supp1}) and (\ref{supp2}).
\newline
4. For non-vanishing temperature $T$ the standard expression for
the phase difference is valid up to correction $\sim
(T/\omega)$.

\section{Acknowledgments}
We are grateful to A. Bondar, M. Danilov,  K. Ter-Martirosyan,
V. Vinokurov, and  M. Vysotsky
for fruitful discussions and interest to this work.  \\

\section{ Appendix A}

\setcounter{equation}{0}
\def\theequation{A.\arabic{equation}}

{\bf 1. General formulas}\\

The aim of this Appendix is to integrate $P_{ij}$ given by
eq.(\ref{Probability}) over the phase space of nuclei $C$ and $D$
and over the energy of the final lepton $l$:
\begin{equation}
\label{B_1} {\cal P}_{i j} = \int P_{i j} \frac{d\p_C}{2E_C}
\frac{d\p_D}{2E_D} \frac{p_l d E_l}{2}.
\end{equation}
In what follows for simplicity we consider an ultra-relativistic
electron:
\begin{equation}\label{ultra-relativistic electron}
 q^0_e=E^0_e,
\end{equation}
where $E_e^0$ and $q_e^0$ are defined by eqs.(\ref{E_e^0}) and
(\ref{q_e^0}).
We assume that the target and the detector contain the same
nuclei:
\begin{equation}\label{MAB sigmaAB}
M_A=M_B=M_C=M_D=M,~\sigma_A=\sigma_B=\sigma.
\end{equation}

The Appendix consists of six parts. In part 1
(eqs.(\ref{B_1})-(\ref{supp-zero})) we make no special assumptions
about the width $\sigma_e$ of the electron wave packet. In parts 2
and 3 (eqs.(\ref{equation for p^0_C})-(\ref{FIN-FIN})) the case of
a relatively broad electron wave packet is considered:
\begin{equation}\label{broad electron wp}
\sigma_e\gg\sigma\frac{p_e}{M}.
\end{equation}
The parts 4 and 5 (eqs.(\ref{condition of small
sigma_e})-(\ref{final})) are devoted to the narrow electron wave
packet:
\begin{equation}\label{narrow electron wp}
\sigma_e\ll\sigma\frac{p_e}{M}.
\end{equation}
Part 6 (eqs.(\ref{plane wave Aj})-(\ref{plane wave Pij
initial})) deals with the case of an electron plane wave:
\begin{equation}\label{electron plane wave}
\sigma_e=0.
\end{equation}

Taking into account eq.(\ref{MAB sigmaAB}) and omitting
pre-exponential factor in eq.(\ref{Probability}), we get:
$$
P_{i j} =
\exp \left( i \frac{\delta m^2_{i j} x_{AB}}{2\omega}\right)
$$
\begin{equation}
\label{B_2}
\exp \left\{-\frac{(E^0_e \e - \p_e)^2}{\sigma_e^2}
           -\frac{(\omega \n + \p_C - E^0_e \e)^2}{\sigma^2}
-\frac{(\omega \n - \p_l - \p_D)^2}{\sigma^2}\right\},
\end{equation}
where $E^0_e$ is defined by eq.(\ref{E_e^0}).

As $\omega$ does not depend
\footnote{We could have defined $\omega$ not by eq.(\ref{omega}) for vertex $B$, but by energy conservation at vertex $A$. However, the oscillation phase still would not depend on momentum of nucleus $C$.
$$\omega(q^0_e,\p_C)=E_C(\p_C)-E_e(q^0_e)-E_A,$$ $q^0_e$ depends on $\p_C$: $$q^0_e=q^0_e(\p_C)$$ in such a way that
$$\frac{d\omega(q^0_e(\p_C),\p_C)}{d\p_C} =\frac{\partial E_C(\p_C)}{\partial \p_C}-\frac{\partial E_e(q^0_e)}{\partial q^0_e}\frac{\partial q^0_e}{\partial \p_C}=0.$$}
on $\p_C$, it is convenient to integrate $P_{ij}$ in eq.(\ref{B_1}) first over $\p_C$ and then over $\p_D$ and $E_l$. For this purpose we introduce function $I_1$:
\begin{equation}
\label{BB_1} {\cal P}_{i j} = \int I_1~ e^{i \frac{\delta m^2_{i
j} x_{AB}}{2 \omega} } \exp \left\{- \frac{(\omega \n - \p_l -
\p_D)^2}{\sigma^2}
    \right\}
\frac{d\p_D}{2E_D} \frac{p_l d E_l}{2},
\end{equation}
where

\begin{equation}\label{integral over dp_C}
I_1\equiv\int \exp \left[-\frac{(E^0_e
\e - \p_e)^2}{\sigma_e^2}-\frac{(\omega \n + \p_C - E^0_e
\e)^2}{\sigma^2}\right] ~p_C~  \sin\theta ~d \theta ~dE_C ~.
\end{equation}
Here $\theta$ is the angle between $\p_C$ and $ E^0_e \e - \omega
\n$, while
\begin{equation}\label{variable substitution}
  \frac{d\p_C}{2E_C}=\pi \sin \theta d \theta p_C d E_C.
\end{equation}
In this Appendix $p_C$ and $p_D$ denote modules of three-vectors
$\p_C$ and $\p_D$, respectively.

Let us integrate over $\theta$ first:
$$I_1=\int  ~dE_C ~\frac{\sigma^2}{|E^0_e\e-\omega \n|}
\exp\left[-\frac{(E^0_e  - p_e)^2}{\sigma_e^2}\right] $$
\begin{equation}
\label{I_1 integrated over the angle}
\left\{ \exp\left[-\frac{(|E^0_e\e-\omega \n|- p_C)^2}
{\sigma^2}\right]
- \exp\left[-\frac{(|E^0_e\e-\omega \n|+ p_C)^2}{\sigma^2}\right]
\right\}.
\end{equation}

It is evident that $I_1$ is not singular at zero angle
$\alpha ({\bf e,n})$ between neutrino and electron. Still
we assume that vectors $\e$ and $\n$ have a substantially
nonzero angle  between them and the module of their difference is
of order of unity, that is
\begin{equation}\label{noncollinearity}
|E^0_e\e-\omega \n|\gg\sigma,
\end{equation}
and thus we may neglect the second term in the curly brackets in
eq.(\ref{I_1 integrated over the angle}).

As was mentioned in the body of the text,
the suppression of oscillations is a consequence of dispersion of
neutrino energy. For $\alpha ({\bf e,n})=0$ and large $\sigma_e$
the suppression length is given by eq.(\ref{supp1}) because
electron energy spread coincides with that of neutrino.

If we consider now the case of very small $\sigma_e$, the
suppression governed by nuclear $\sigma$ becomes dominating.
Let us assume, for simplicity, that $m_e=0$ and nucleus $C$ does not
interact with the crystal. For zero angle production of neutrino
the momentum of the nucleus $C$ is equal to zero up to $\sigma$, while its energy does not exceed the value of the order
of $\sigma^2 /M$, and hence neutrino energy equals $E_e$ up to
this value. We omit the derivation noting only that the suppression length in this case is very large:
\begin{equation}
\label{supp-zero}
L_{ij}^{sup} \sim L_{ij}^{osc} ~\frac{p_e M}{\sigma^2}.
\end{equation}

{\bf 2. Large $\sigma_e$, integration over $p_C$}. \\

Now we consider the case of relatively broad electron wave packet. We assume that the first exponent in the curly brackets is much sharper than one which is out of the brackets, and cuts out the effective region of integration. The condition under which one can make such assumptions will be obtained further (see eq.(\ref{large sigmae})).

The dominant exponent (the first in the curly brackets) has a
maximum when
\begin{equation}\label{equation for p^0_C}
f(E_C)\equiv(|E^0_e(E_C)\e-\omega \n|- p_C(E_C))^2=0.
\end{equation}
Note that $E_e^0$ is "chosen" by global conservation of energy
(eq.(\ref{E_e^0})), and hence depends on $E_C$.

Let us denote the solution of this equation as $E^0_C$. It has
the sense of the most probable value of $E_C$.
Evidently $p_C^0 \equiv \sqrt {(E_C^0)^2 - M^2}$. For non-vanishing
values of the angle $\alpha ({\bf e}, {\bf n})$ defined above
$p_C^0 \approx |\p_e-\omega \n| \sim p_e$.
Here and in what follows the sign
"$\sim$" means "is of the order of magnitude".

We expand $f(E_C)$ up to the third order in $(E_C-E^0_C)$:
\begin{equation}\label{expantion of the exponent}
f(E_C)= \frac{1}{2}(E_C-E^0_C)^2 \frac{d^2f}{dE^2_C}(E^0_C)
+\frac{1}{6}(E_C-E^0_C)^3 \frac{d^3f}{dE^3_C}(E^0_C).
\end{equation}

Taking  into account that $p^0_C\sim p_e$ we estimate
the derivatives:

\begin{equation}\label{estimation of derivatives}
\frac{1}{2}\frac{d^2f(E^0_C)}{dE^2_C}\simeq (\frac{M}{p^0_C})^2\sim\frac{M^2}{p^2_e} ~,~~~~~~~~
\frac{1}{6}
    \frac{d^3f(E^0_C)}{dE^3_C}\simeq - \frac{M^3}{(p^0_C)^4} \sim
    -\frac{M^3}{p^4_e}~.
\end{equation}
Now we can see that the
effective width of the exponent in the curly brackets in (\ref{I_1
integrated over the angle}) is $\sigma p_e/M$ while the width
of the first exponent in (\ref{I_1 integrated over the angle}) is
$\sigma_e$. Thus the electron wave packet may be considered as
a broad one if
\begin{equation}\label{large sigmae}
\sigma_e\gg\sigma\frac{p_e}{M}.
\end{equation}

From eqs.(\ref{I_1 integrated over the angle}), (\ref{expantion of
the exponent}) and (\ref{estimation of derivatives}) we have
$$ I_1 \simeq \int  ~dE_C ~\frac{\sigma^2}{|E^0_e\e-\omega \n|}
\exp\left[\frac{M^3 (E_C-E^0_C)^3}{(p^0_C)^4\sigma^2}-\frac{(E_C
-E^0_C)^2}{\sigma_e^2}\right]$$
$$ \exp\left[-\frac{(E^0_C+\omega-E_A-p_e)^2}{\sigma^2_e}
(1-\left(\frac{\sigma
p^0_C}{\sigma_e M}\right)^2) \right]$$
\begin{equation}\label{prefinal I_1}
\exp\left[-\frac{M^2\left(E_C-E^0_C+(E^0_C+\omega-E_A-p_e)
\left(\frac{\sigma
p^0_C}{\sigma_eM}\right)^2\right)^2}{(\sigma p^0_C)^2}\right].
\end{equation}

Note that the integrand in  $I_1$ is not vanishingly small if
\begin{equation}\label{estimation 0.0}
\left|E^0_C-E_A-p_e+\omega \right|\lesssim \sigma_e,
\end{equation}
\begin{equation}\label{estimation 0.1}
\left| E_C-E^0_C+(E^0_C+\omega-E_A-p_e)\left(\frac{\sigma
p^0_C}{\sigma_eM}\right)^2
\right| \lesssim \frac{\sigma p^0_C}{M}
\end{equation}
and thus taking into account eq.(\ref{large sigmae}) we may
estimate the essential range of $E_C-E^0_C$:
\begin{equation}\label{estimation 0.2}
|E_C-E^0_C|\lesssim \frac{\sigma p^0_C}{M}.
\end{equation}

This means that the  exponent in the first line of eq.(\ref{estimation 0.0})
may be omitted. Furthermore, the small corrections
in the second and third lines of eq.(\ref{estimation 0.0}) proportional
to $\left(\frac{\sigma p^0_C}{\sigma_e M}\right)^2$  are negligible.
Using this we obtain
\begin{equation}
\label{I_1 once more time}
I_1 \simeq \int  dE_C \frac{\sigma^2}{|E^0_e\e-\omega \n|}
\exp\left[-\frac{(E^0_C+\omega-E_A-p_e)^2}{\sigma^2_e}
-\frac{M^2(E_C-E^0_C)^2}{(\sigma p^0_C)^2}\right]
\end{equation}
and, after integration:

\begin{equation}\label{final I_1}
I_1= \frac{\sigma^3}{M}
\exp\left[-\frac{(E^0_C -p_e-E_A+\omega)^2}{\sigma_e^2}\right]~.
\end{equation}   \\

{\bf 3. Large $\sigma_e$; integration over $\p_D$ and $E_l$.} \\

First we integrate  eq.(\ref{BB_1}) over the angle
between $\p_D$ and $\omega \n - \p_l$ and obtain:
$$
\mathcal{P}_{ij}= \int dE_DdE_l \frac{p_l}{~|\omega \n - \p_l|}
 e^{i\frac{\delta m^2_{i j} x_{AB}}{2 \omega} }
 \exp\left[-\frac{(E^0_C
-p_e-E_A+\omega)^2}{\sigma_e^2}\right]
$$
\begin{equation}\label{I_2 integrated over the angle}
 \frac{\sigma^5}{M}
\left\{\exp\left[-\frac{(|\omega \n - \p_l|- p_D)^2}{\sigma^2}\right]
- \exp\left[-\frac{(|\omega \n - \p_l|+ p_D)^2}{\sigma^2}\right]\right\}.
\end{equation}
Assuming that the angle between vectors $\n$ and ${\mathbf l}
\equiv\p_l/p_l$ satisfies the inequality
\begin{equation}\label{inequality for alpha(nl)}
\alpha(\n,{\mathbf l})\gg\frac{\sigma}{p_e},
\end{equation}
we disregard the second exponent in the curly brackets in
eq.(\ref{I_2 integrated over the angle}).

To integrate over $E_D$ and $E_l$ we introduce
\begin{equation}\label{F}
F(E_D,E_l)\equiv \frac{1}{\sigma_e^2}(E^0_C -p_e-E_A+\omega)^2+
\frac{1}{\sigma^2}(|\omega\n-p_l{\bf l}|-p_D)^2.
\end{equation}

Expanding all functions of $E_D$ and $E_l$ in the exponents in
eq.(\ref{I_2 integrated over the angle}) around the most probable
values $E^0_D$ and $E_l^0$, which are solution of the equation
\begin{equation}\label{equation for E^0_D}
F(E_D,E_l)=0,
\end{equation}
we obtain from eq.(\ref{I_2 integrated over the angle})

$$
\mathcal{P}_{ij}=\frac{p_l^0}{|p_l^0{\bf l}-E_{\nu} \n|}
\frac{\sigma^5}{M} e^{i\frac{\delta m_{ij}^2x_{AB}}{2E_{\nu}}}\int
dE_DdE_le^{-i\frac{\delta m_{ij}^2x_{AB}}{2E^2_{\nu}}(\Delta
E_D+\Delta E_l)}
$$
\begin{equation}\label{large sigma_e Pij prefinal}
\exp\left[-\frac{M^2}{(\sigma p_D^0)^2}
(\Delta E_D)^2-2(\frac{1}{\sigma_e^2}-\frac{\lambda M}
{\sigma^2p_D^0})\Delta E_D\Delta E_l-
(\frac{1}{\sigma_e^2}+\frac{\lambda^2}{\sigma^2})
(\Delta E_l)^2\right],
\end{equation}\\

where $\Delta E_D\equiv E_D-E_D^0$, $\Delta E_l\equiv E_l-E_l^0$,
\begin{equation}\label{lambda}
\lambda\equiv\frac{(E_{\nu}\n-p_l^0{\bf
l})(\n-\frac{E_l^0}{p_l^0}{\bf l})}{|E_{\nu}\n-p_l^0{\bf l}|},
\end{equation}
$p_l^0\equiv\sqrt{(E_l^0)^2-m_l^2}$, $m_l$ stands for the mass of
the final lepton, $p_D^0\equiv\sqrt{(E_D^0)^2-M^2}$ and
$E_{\nu}\equiv E^0_D+E^0_l-E_B$.

Taking into account that $\lambda\sim 1$ we may integrate over
$E_D$ and $E_l$ in eq.(\ref{large sigma_e Pij prefinal}):

\begin{equation}\label{FIN-FIN}
{\cal P}_{i j} = \sigma^6 \frac{p_e \sigma_e}{M^2}
\exp\left[i \frac{\delta m^2_{i j} x_{AB}}{2 E_{\nu}}
-\left(\frac{\delta m^2_{i j} x_{AB}}{2
E_{\nu}} \frac{\sigma_e}{2  E_{\nu}} \right)^2\right],
\end{equation}\\
where we use eqs.(\ref{F}) and (\ref{equation for E^0_D}) which
define $E_D^0$ and $E_l^0$.\\

{\bf 4. Small $\sigma_e$; integration over $p_C$}. \\

Let us now consider the case of small $\sigma_e$, that is
\begin{equation}\label{condition of small sigma_e}
\sigma_e\ll\frac{p_e}{M}\sigma.
\end{equation}
The integration will be mainly carried out as for the case of
large $\sigma_e$.

The first exponent in eq.(\ref{I_1 integrated over the angle})
has a sharp maximum when
\begin{equation}\label{small sigma_e-equation for p^0_C}
E_C=E_C^0\equiv p_e+E_A+E_B-E_l-E_D.
\end{equation}
As before, $E_C^0$ has the sense of the most probable value of $E_C$. We would like to emphasize that the definition of $E_C^0$ here is different
from one which is for the case of large $\sigma_e$ (see eqs.(\ref{equation for p^0_C}) and (\ref{small sigma_e-equation for p^0_C})). To estimate the contribution of the first exponent in the curly brackets in
eq.(\ref{I_1 integrated over the angle}) it is convenient to use the
function $f(E_C)$ defined in eq.(\ref{equation for p^0_C}).

Expanding $f(E_C)$ near $E_C^0$ to the first order of $\Delta
E_C\equiv E_C-E_C^0$ we obtain for $I_1$ from eq.(\ref{I_1
integrated over the angle}):
\begin{equation}\label{Small sigma_e I_1}
I_1= \frac{\sigma^2}{|p_e\e-\omega \n|}\int  ~dE_C
\exp(-\frac{(\Delta E_C)^2}{\sigma_e^2})
 \exp\left[-\frac{f(E_C^0)+\frac{df}{dE_C}(E_C^0)\Delta
E_C}{\sigma^2}\right].
\end{equation}
Since
\begin{equation}\label{Neglible of df/dE_C}
\frac{df}{dE_C}(E_C^0)\sim\sigma\frac{M}{p_e}~,
\end{equation}
then the term containing derivative in eq.(\ref{Small sigma_e
I_1}) is negligible,
\begin{equation}\label{terms with df/dE_C}
\frac{1}{\sigma^2}\frac{df}{dE_C}(E_C^0)\Delta
E_C\sim\frac{\sigma_e}{\sigma}\frac{M}{p_e}\ll 1,
\end{equation}
and can be omitted. Granting this we integrate in (\ref{Small
sigma_e I_1}) and obtain
\begin{equation}\label{Small sigma_e I_1 final}
I_1= \frac{\sigma^2 \sigma_e}{|p_e\e-\omega \n|}
\exp\left[-\frac{(|p_e\e-\omega\n|-p_C^0)^2}{\sigma^2}\right],
\end{equation}
where $p_C^0\equiv\sqrt{(E_C^0)^2-M^2}$. Note that exponents in
eqs.(\ref{final I_1}) and (\ref{Small sigma_e I_1 final}) are
different. The value of $\sigma_e$ determines which exponent in
eq.(\ref{I_1 integrated over the angle}) is more narrow and cuts out the effective region of integration. The wider exponent remains after integration in (\ref{I_1 integrated over the angle}).\\

{\bf 5. Small $\sigma_e$; integration over $\p_D$ and $E_l$.} \\

Similarly to the case of large $\sigma_e$ integration in
(\ref{BB_1}) over the angle between $\p_D$ and $\omega \n - \p_l$
gives:
$$
\mathcal{P}_{ij}=\int dE_ldE_D \frac{p_{l} \sigma^4
\sigma_e}{|p_e\e-\omega\n|\cdot|\p_l-\omega\n|}
$$
\begin{equation}\label{small sigma_e I_2}
 \exp\left[i\frac{\delta m_{ij}^2x_{AB}}{2\omega}
 -\frac{(|p_e\e-\omega\n|-p_C^0)^2}{\sigma^2}
 -\frac{(|\p_l-\omega\n|-p_D)^2}{\sigma^2}\right],
\end{equation}
where we take into account the assumption (\ref{inequality for
alpha(nl)}).

We integrate over $E_D$ and $E_l$ analogously to the case of large
$\sigma_e$. Introducing
\begin{equation}\label{f2}
G(E_D,E_l)\equiv
(|p_e\e-\omega\n|-p_C^0)^2+(|\p_l-\omega\n|-p_D)^2
\end{equation}
and expanding all functions of $E_D$ and $E_l$ in exponents in
eq.(\ref{small sigma_e I_2}) near the most probable values $E_D^0$
and $E_l^0$, which are defined by equation
\begin{equation}\label{small sigma_e E_D^0}
G(E_D,E_l)=0,
\end{equation}
we obtain
$$
\mathcal{P}_{ij}\simeq\frac{p_l^0 \sigma^4
\sigma_e}{|p_e\e-E_{\nu}\n|\cdot|p_l^0{\bf l}
-E_{\nu}\n|}e^{i\frac{\delta m_{ij}^2x_{AB}}{2E_{\nu}}}\int
dE_DdE_le^{-i\frac{\delta m_{ij}^2x_{AB}}{2(E_{\nu})^2}(\Delta
E_D+\Delta E_l)}
$$
\begin{equation}\label{small sigma_e I_2 prefinal}
\exp\left\{-\frac{M^2}{\sigma^2}
\left[\frac{(\Delta E_D)^2}{(p_D^0)^2}+
\frac{(\Delta E_D)^2}{p_e^2(\e-\n)^2}
+\frac{2 \Delta E_D\Delta E_l}{p_e^2(\e-\n)^2}
+\frac{(\Delta E_l)^2}{p_e^2(\e-\n)^2}\right]\right\},
\end{equation}
where $\Delta E_D\equiv E_D-E_D^0$, $\Delta E_l\equiv E_l-E_l^0$,
$p_l^0\equiv\sqrt{(E_l^0)^2-m_l^2}$,
$p_D^0\equiv\sqrt{(E_D^0)^2-M^2}$ and $E_{\nu}\equiv
E^0_D+E^0_l-E_B$.

After integration in (\ref{small sigma_e I_2 prefinal}) taking
into account eqs.(\ref{f2}) and (\ref{small sigma_e E_D^0}) we
finally obtain
\begin{equation}\label{final}
\mathcal{P}_{ij}\simeq\sigma^6 \frac{p_e \sigma_e}{M^2}~
\exp\left[i\frac{\delta m_{ij}^2x_{AB}}{2E_{\nu}}~
-\left(\frac{\delta
m_{ij}^2x_{AB}}{2E_{\nu}}~~
\frac{p_e|\e-\n|\sigma}{2ME_{\nu}}
\right)^2\right].
\end{equation}

{\bf 6. The case of electron plane wave.}\\

In this part of Appendix we would like to discuss separately
the case of the incident electron with definite momentum $\p_e$.
As is well known in the limil $\sigma_e \rightarrow 0$ Gaussian
$\frac{1}{(2\pi)^{3/2}\sigma_e^3}
exp\left(\frac{(\q_e-\p_e)^2}{\sigma_e^2} \right) \rightarrow \delta^3(\q_e-\p_e)$.
By substituting it in eq.(\ref{final A_j}) $K_e(\q_e-\p_e)=$$\delta(\q_e-\p_e)$ we immediately obtain
the following expression for the amplitude
\begin{equation}\label{plane wave Aj} A_j =
e^{ik_j x_{AB}} e^{i\p_e(\x_A-\x_e)}
\exp(-\frac{({\bf q}_A)^2}{2\sigma^2}) \exp(-\frac{({\bf
q}_B)^2}{2\sigma^2}) \delta (E_{in} - E_{fin}),
\end{equation}
where $\q_A=\p_C+\omega\n-\p_e$, $\q_B=\p_l+\p_D-\omega\n$, and
as previously we assume the nuclei to be in the oscillator-like
potential.

We use
\begin{equation}\label{delta squared}
\left(\delta (E_{in} - E_{fin})\right)^2=
T ~\delta (E_{in} - E_{fin}),
\end{equation}
where $T$ is a duration of the experiment, to square the
delta-function in eq.(\ref{plane wave Aj}), and obtain the
probability normalized per unit of time:

\begin{equation}
\label{plane wave Pij initial}
P_{ij}= \sigma^2 ~
\delta (E_{in} - E_{fin})
\exp\left[{
i\frac{\delta m_{ij}^2}{2\omega}x_{AB}
-\frac{(\omega\bf{n}\mit+\bf{p}\mit_C-\p_e)^2}{\sigma^2}}
{-\frac{(\bf{p}\mit_l+\bf{p}\mit_D-\omega\bf{n}\mit)^2}
{\sigma^2}}\right].
\end{equation}
We may integrate eq.(\ref{plane wave Pij initial}) over
$\p_C$, $\p_D$ and $E_l$ exactly as in the case of small $\sigma_e$
(see eqs.(\ref{small sigma_e I_2})-(\ref{small sigma_e I_2 prefinal})).
This procedure results in eq.(\ref{final}).

\end{document}